\documentclass[11pt]{article}
\pdfoutput=1

\usepackage{jheppub}
\usepackage{bigints}
\usepackage{rotating}

\usepackage{tabularx}

\usepackage{tikz}
\usepackage{fancybox}
 
\usepackage{tikz-cd}
\usepackage{ae}
\usepackage{url}

\usepackage{subfig}

\usepackage{tabularx}

\usepackage{graphicx}
\usepackage{dsfont}
\usepackage{setspace}
\usepackage{amsfonts,amsmath,amsthm,amssymb,amsbsy}
\usepackage{longtable}
\usepackage{array}
\usepackage{color}
\usepackage{needspace}
\usepackage[numbers]{natbib}
\usepackage{colortbl}

\usepackage{enumitem}

\usepackage{fancyhdr}

\usepackage{tabularx}

\usepackage{graphicx}
\usepackage{dsfont}
\usepackage{setspace}
\usepackage{amsfonts,amsmath,amsthm,amssymb,amsbsy}
\usepackage{longtable}
\usepackage[latin1]{inputenc}
\usepackage{array}
\usepackage{color}
\usepackage{needspace}
\usepackage[numbers]{natbib}
\usepackage{colortbl}

\usepackage{shuffle}

\usepackage{xcolor}
\definecolor{colourO}{HTML}{4FEEBB}
\definecolor{colourL}{HTML}{5DB197}
\definecolor{colourG}{HTML}{3E7CEF}
\definecolor{colourK}{HTML}{3535E3}

\usepackage{mathtools}

\usepackage{enumitem}

\usepackage{fancyhdr}

\usepackage{booktabs}
\usepackage{tcolorbox}
\usepackage{tabularx}

\usepackage{tensor}
\usepackage{mathtools}
\usepackage{blkarray, bigstrut}
\usepackage{soul}

\usepackage{physics}
\usepackage{empheq}

\title{The hypersimplex canonical forms and the momentum amplituhedron-like logarithmic forms}

\author[1]{Tomasz \L ukowski,}\emailAdd{t.lukowski@herts.ac.uk}
\author[1]{and Jonah Stalknecht}\emailAdd{j.stalknecht@herts.ac.uk}

\affiliation[1]{Department of Physics, Astronomy and Mathematics, \\ University of Hertfordshire, \\  Hatfield, Hertfordshire, AL10 9AB, United Kingdom}

\abstract{In this paper we provide a formula for the canonical differential form of the hypersimplex $\Delta_{k,n}$ for all $n$ and $k$. We also study  the generalization of the momentum amplituhedron $\mathcal{M}_{n,k}$ to $m=2$, and we conclude that the existing definition does not possess the desired properties. Nevertheless, we find interesting momentum amplituhedron-like logarithmic differential forms in the $m=2$ version of the spinor helicity space, that have the same singularity structure as the hypersimplex canonical forms.}
\begin{document}

\maketitle


\section{Introduction}
Geometry has always played an essential role in physics, and it continues to be crucial in many recently developed branches of theoretical and high-energy physics. In recent years, this statement has been supported by the introduction of positive geometries \cite{Arkani-Hamed:2017tmz} that encode a variety of observables in Quantum Field Theories \cite{Arkani-Hamed:2013jha, Arkani-Hamed:2017mur, Damgaard:2019ztj, Arkani-Hamed:2020blm}, and beyond \cite{Arkani-Hamed:2018ign, Arkani-Hamed:2017fdk, Arkani-Hamed:2019mrd}, see \cite{Ferro:2020ygk} for a comprehensive review. These recent advances have also renewed the interest in well-established and very well-studied geometric objects, allowing us to look at them in a completely new way. One essential new ingredient introduced by positive geometries is that to every convex polytope, one can associate a meromorphic differential form with the property that it is singular on all boundaries of the polytope, and the divergence is logarithmic. Moreover, when each boundary is approached, an appropriately defined residue operation allows one to find the differential form of the boundary with the same properties. This process can be repeated and eventually one arrives at a zero-dimensional boundary with a trivial 0-form equal $\pm 1$. Such canonical forms can be found for every convex polytope and for more complicated ``convex'' shapes in Grassmannian spaces, like the amplituhedron \cite{Arkani-Hamed:2013jha} or the momentum amplituhedron \cite{Damgaard:2019ztj}. Many well-known convex polytopes made their recent appearance in physics in the context of positive geometries, the primary example given by the associahedron featured in the bi-adjoint $\phi^3$ scalar field theory \cite{Arkani-Hamed:2017mur} or, more generally, generalized permutahedra discussed in \cite{He:2020onr}. More recently, another well-known polytope, the hypersimplex $\Delta_{k,n}$, also has become relevant in the positive geometry story. It was shown in \cite{Lukowski:2020dpn} that a particular class of hypersimplex subdivisions are in one-to-one correspondence with the triangulations of the amplituhedron $\mathcal{A}_{n,k}^{(2)}$, which is a prototypical example of a positive geometry. Moreover, it was conjectured  that its spinor helicity cousin, $\mathcal{M}_{n,k}^{(2)}$, which is a generalization of the momentum amplituhedron $\mathcal{M}_{n,k}$ \cite{Damgaard:2019ztj}, shares many properties with the hypersimplex. This paper focuses on the latter statement and tries to verify whether it is correct. To this extent, we start by treating the hypersimplex as a positive geometry and finding its canonical differential form. In particular, the hypersimplex $\Delta_{k,n}$ can be defined as the image of the positive Grassmannian through the (algebraic) moment map. Using this fact, we find a simple expression for the hypersimplex canonical form, which can be obtained by summing push-forwards of canonical forms of particular cells in the positive Grassmannian $G_+(k,n)$. The momentum amplituhedron $\mathcal{M}_{n,k}^{(2)}$ has also been defined as the image of the same positive Grassmannian using a linear map $\Phi_{\Lambda,\widetilde\Lambda}$ \cite{Lukowski:2020dpn}, which we will define in the main text. After taking the same collection of positroid cells in the positive Grassmannian, and summing their push-forwards through the $\Phi_{\Lambda,\widetilde\Lambda}$ map, we find a simple logarithmic differential form in spinor helicity space, that has the same singularity structure as the hypersimplex canonical form. However, it is not the canonical form of $\mathcal{M}_{n,k}^{(2)}$. Moreover, we show that $\mathcal{M}_{n,k}^{(2)}$ does not possess the desired properties conjectured in \cite{Lukowski:2020dpn}.

The paper is organized as follows: in Section \ref{sec:hypersimplex} we recall the definition of hypersimplex, describe its boundary structure and define positroid triangulations. We also provide a previously unknown formula for its canonical differential form. In section \ref{sec:mom} we recall the definition of the momentum amplituhedron $\mathcal{M}_{n,k}^{(2)}$ introduced in \cite{Lukowski:2020dpn} and find a logarithmic differential form defined on the $m=2$ version of the spinor helicity space, that has the same singularity structure as the hypersimplex canonical form. We also comment on the validity of the conjectures in section 12 of \cite{Lukowski:2020dpn}. We end the paper with a summary and outlook, and an appendix containing the definitions of positive geometries and push-forwards.


\section{Hypersimplex}
\label{sec:hypersimplex}
The hypersimplices $\Delta_{k,n}$ form a two-parameter family of convex polytopes that appears in various algebraic and geometric contexts. In particular, they have been used to classify points in the Grassmannian $G(k,n)$ by studying their images through the moment map \cite{GGMS}. This naturally leads to a notion of matroid polytopes and matroid subdivisions \cite{Kapranov, Lafforgue,Speyer}, which are in turn related to the tropical Grassmmanian \cite{tropgrass,Kapranov,Dressian}. When the Grassmannian $G(k,n)$ is replaced by its positive part $G_+(k,n)$, the moment map image of $G_+(k,n)$ is still the hypersimplex $\Delta_{k,n}$, and one can use it to study positroid polytopes \cite{tsukerman_williams}, positroid subdivisions \cite{Lukowski:2020dpn, Arkani-Hamed:2020cig,Early:2019zyi} and their relation to the positive tropical Grassmannian \cite{troppos}.  In this paper we look at the hypersimplex $\Delta_{k,n}$ from the point of view of positive geometries\footnote{For an introduction on positive geometries, we refer the reader to \cite{Arkani-Hamed:2017tmz}, we also collect some basic information in appendix \ref{app:positive}.
}. As the main result of this section, we provide an explicit expression for the canonical differential form for $\Delta_{k,n}$ for all $n$ and $k$. 
\subsection{Definitions}
We denote by $e_i$ the standard basis vectors in $\mathbb{R}^n$. The hypersimplex $\Delta_{k,n}$ is then defined as the convex hull of the indicator vectors $e_I=\sum_{i\in I}e_i$ where $I$ is a $k$-element subset of $[n]\equiv\{1,2,\ldots,n\}$. Since for all $x=(x_1,\ldots,x_n)\in\Delta_{k,n}$ we have $x_1+\ldots+x_n=k$, the hypersimplex $\Delta_{k,n}$ lives in an $(n-1)$-dimensional affine subspace inside $\mathbb{R}^n$. Moreover, the hypersimplex $\Delta_{k,n}$ is identical to the hypersimplex $\Delta_{n-k,n}$ after the replacement $I\leftrightarrow [n]\setminus I$. We refer to this symmetry as a {\it parity symmetry}.

Equivalently, the hypersimplex $\Delta_{k,n}$ can be defined as the image of the positive Grassmannian $G_+(k,n)$ through the moment map \cite{GGMS}. For a given $n$ and $0\leq k\leq n$, the {\it Grassmannian} $G(k,n)$ is the space of all $k$-dimensional subspaces of $\mathbb{R}^n$. Each element of $G(k,n)$ can be viewed as a maximal rank $k\times n$ matrix modulo $GL(k)$ transformations, which gives a basis for the $k$-dimensional space. We denote by $\binom{[n]}{k}$ the set of all $k$-element subsets of $[n]$. Then for $I\in \binom{[n]}{k}$, we define $p_I(C)$ to be the maximal minor formed of columns of $C$ labelled by elements of $I$. We call these variables the {\it Pl\"{u}cker variables}, and they are defined up to an overall rescaling by a non-zero constant. The {\it positive Grassmannian} $G_+(k,n)$ is the set of all elements $C\in G(k,n)$ for which $p_I(C)\geq0$ for all $I\in \binom{[n]}{k}$. Finally, we define the {\it moment map} 
\begin{equation}
\mu:G(k,n)\to\mathbb{R}^n\,,
\end{equation}
as 
\begin{equation}
\mu(C)=\frac{\sum_I |p_I(C)|^2e_I}{\sum_I |p_I(C)|^2}\,.
\end{equation}
Then, the hypersimplex is the image of the (positive) Grassmannian 
\begin{equation}
\Delta_{k,n}=\mu(G(k,n))=\mu(G_+(k,n))\,.
\end{equation}
If we restrict our attention to the positive Grassmannian $G_+(k,n)$, we can instead use the algebraic moment map \cite{Sottile}
\begin{equation}
\tilde\mu(C)=\frac{\sum_I p_I(C)e_I}{\sum_I p_I(C)}\,,
\end{equation}
which will significantly simplify our calculations in the following. Most importantly, we have 
\begin{equation}
\Delta_{k,n}=\tilde\mu(G_+(k,n))\,,
\end{equation}
see \cite{Lukowski:2020dpn} for more details.

An important fact we will use later is that the positive Grassmannian $G_+(k,n)$ has a natural decomposition into cells of all dimensions \cite{Postnikov:2006kva}. For a subset $M\subset \binom{[n]}{k}$, we denote by $S_M$ the subset of all elements in the positive Grassmannian $G_+(k,n)$ such that its Pl\"ucker variables are positive, $p_I>0$, for $I\in M$, and they vanish, $p_I=0$, for $I\not\in M$. If $S_M\neq \emptyset$ then we call $S_M$ a {\it positroid cell}. Positroid cells can be labelled by various combinatorial objects, most importantly by affine permutations $\pi$ on $[n]$, see \cite{Postnikov:2006kva} for a review of this labelling. From now on we will use $S_\pi$ instead of $S_M$ to label positroid cells of positive Grassmannian.

In the following, we will adopt the notation from \cite{Lukowski:2020dpn}. The image of the positroid cell $S_\pi$ through the algebraic moment map $\tilde\mu$ is called a {\it positroid polytope}, and we denote it by $\Gamma_\pi=\tilde\mu(S_\pi)$. We will be interested in a particular type of positroid polytopes: if the dimension of $\Gamma_\pi$ is $n-1$ and $\tilde\mu$ is injective on $S_\pi$ then we call $\Gamma_\pi$ a {\it generalized triangle}. We will use generalized triangles to define positroid triangulations of the hypersimplex $\Delta_{k,n}$, which will allow us to find its canonical differential form $\omega_{k,n}$. One important property of this differential form is that it is logarithmically divergent on all boundaries of the hypersimplex $\Delta_{k,n}$. These boundaries are also positroid polytopes, of dimension $n-2$, and can be described using the underlying cell decomposition of the positive Grassmannian $G_+(k,n)$. In particular, for $1<k<n-1$, there are exactly $2n$ boundaries of the hypersimplex $\Delta_{k,n}$, and they come in two types: $x_i=0$ or $x_i=1$, for $i=1,\ldots,n$. In the former case, they are images of positroid cells $S_\pi$ with $\dim S_\pi= (k-1)(n-k)$, and the positroid polytope $\Gamma_\pi$ is identical with the hypersimplex $\Delta_{k-1,n-1}$. In the latter case, we find positroid cells $S_\pi$ with $\dim S_\pi= k(n-k-1)$, and $\Gamma_\pi$ is identical with the hypersimplex $\Delta_{k,n-1}$. The exceptional cases are for $k=1$ or $k=n-1$ when the hypersimplices $\Delta_{1,n}$ and $\Delta_{n-1,n}$ are just simplices, with only one type of boundaries: $x_i=0$ for $k=1$ and $x_i=1$ for $k=n-1$. In all these cases, the permutations corresponding to boundary positroid polytopes can be found using the package \texttt{amplituhedronBoudaries} \cite{Lukowski:2020bya}. The package also provides an easy way to find the complete boundary stratification of the hypersimplex $\Delta_{k,n}$.

\subsection{Hypersimplex canonical forms}
\label{sec:hyp-form}
We are now ready to explain how to find the canonical differential form $\omega_{k,n}$ for the hypersimplex $\Delta_{k,n}$. We will use the fact that all hypersimplices can be subdivided using a collection of generalized triangles that are non-overlapping and are dense in $\Delta_{k,n}$. In such a case, the canonical differential form $\omega_{k,n}$ can be found as a sum of push-forwards through the algebraic moment map $\tilde\mu$ of the canonical forms of the corresponding positroid cells in the positive Grassmannian $G_+(k,n)$. More specifically, if $\mathcal{T}=\{\pi_1,\ldots,\pi_p\}$, with $S_{\pi_i}\subset G_+(k,n)$ a positroid cell for $i=1,\ldots,p$, is a collection of affine permutations for which $\{\Gamma_{\pi_1},\ldots,\Gamma_{\pi_p}\}$ is a positroid triangulation of $\Delta_{k,n}$, then
\begin{equation}
\omega_{k,n}=\sum_{\pi\in \mathcal{T}} \tilde\mu_*\, \omega_\pi\,,
\end{equation}   
where $\omega_\pi$ is the canonical form of the positroid cell $S_\pi$, and $\tilde\mu_*$ indicates the push-forward through $\tilde\mu$ defined in Appendix \ref{app:positive}.

As already mentioned, the hypersimplex $\Delta_{k,n}$ reduces to a simplex for $k=1$ or $k=n-1$. In these cases no triangulation is required since the algebraic moment map is already injective, and we can take the push-forward of the top form on the positive Grassmannian $G_+(1,n)$ or $G_+(n-1,n)$. A simple calculation leads to the following canonical differential forms
\begin{align}
\omega_{1,n}&=\dd\log\left(\frac{x_2}{x_1}\right)\wedge\dots\wedge\dd\log\left(\frac{x_n}{x_1}\right),\\
\omega_{n-1,n}&=\dd\log\left(\frac{1-x_2}{1-x_1}\right)\wedge\dots\wedge\dd\log\left(\frac{1-x_n}{1-x_1}\right).
\end{align}
These are just canonical differential forms on the projective space $\mathbb{P}^{n-1}$, with homogeneous coordinates $(x_1,\ldots,x_n)$ in the first case and $(y_1,\ldots,y_n)=(1-x_1,\ldots,1-x_n)$ in the second case.  

\begin{figure}
\begin{center}
\includegraphics[scale=0.5]{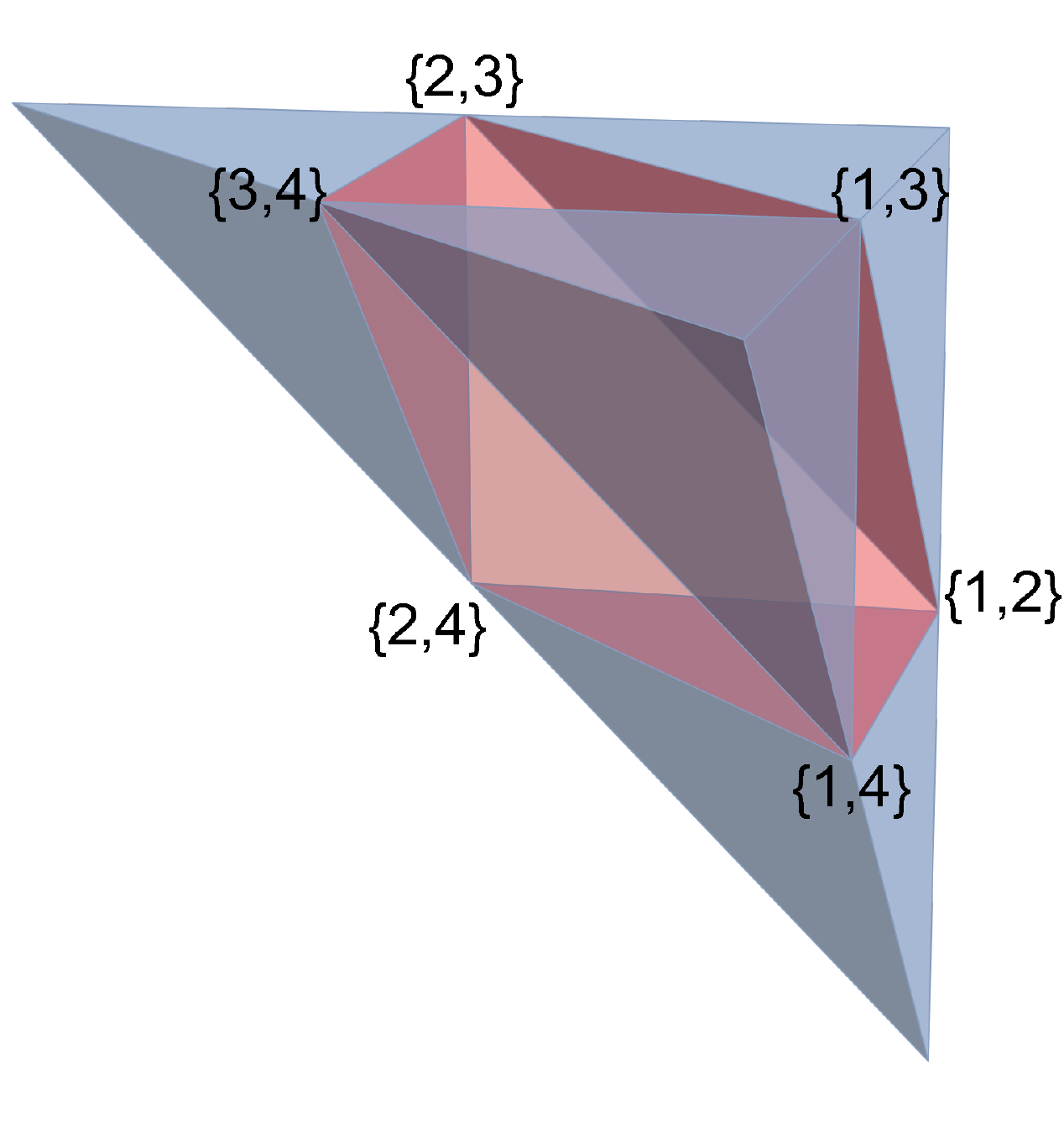}
\end{center}
\caption{Inclusion-exclusion triangulation of $\Delta_{2,4}$ using three-dimensional simplices.}
\label{fig:Delta24}
\end{figure}

For $1<k<n-1$, the algebraic moment map $\tilde\mu$ is not injective anymore, and the image of the positive Grassmannian through $\tilde\mu$ covers the hypersimplex $\Delta_{k,n}$ infinitely many times. To find the canonical form $\omega_{k,n}$ we need to divide the hypersimplex into smaller non-overlapping pieces for which the algebraic moment map is injective, namely generalized triangles, such that their union is dense in $\Delta_{k,n}$. Such subdivisions are called {\it positroid triangulations}, and have been extensively studied in \cite{Lukowski:2020dpn}, where they were related to subdivisions of the amplituhedron \cite{Arkani-Hamed:2013jha}, and to the positive tropical Grassmannian \cite{troppos}. For our purposes, we need to find a single positroid triangulation for a given hypersimplex $\Delta_{k,n}$. There are various ways to find such triangulations: for example using the amplituhedron and T-duality \cite{Lukowski:2020dpn}, or using blade arrangements \cite{Early:2019zyi}. In the simplest non-trivial example, $\Delta_{2,4}$, one finds two positroid triangulations:
\begin{itemize}
\item positroid polytope $\Gamma_{\{3,5,4,6\}}$ with vertices $\{e_{\{1,2\}},e_{\{1,3\}},e_{\{1,4\}},e_{\{2,3\}},e_{\{2,4\}}\}$ and  positroid polytope $\Gamma_{\{2,4,5,7\}}$ with vertices $\{e_{\{1,3\}},e_{\{1,4\}},e_{\{2,3\}},e_{\{2,4\}},e_{\{3,4\}}\}$, or
\item positroid polytope $\Gamma_{\{4,3,5,6\}}$ with vertices $\{e_{\{1,2\}},e_{\{1,3\}},e_{\{1,4\}},e_{\{2,4\}},e_{\{3,4\}}\}$ and  positroid polytope $\Gamma_{\{3,4,6,5\}}$ with vertices $\{e_{\{1,2\}},e_{\{1,3\}},e_{\{2,3\}},e_{\{2,4\}},e_{\{3,4\}}\}$
\end{itemize}
where we explicitly specified the affine permutations labelling cells in $G_+(2,4)$. 
Each of these polytopes is the image of a positroid cell $S_\pi$ in the positive Grassmannian $G_+(2,4)$, and the algebraic moment map $\tilde\mu$ is injective on all of them. This allows us to invert $\tilde\mu$ on these cells, and to find the push-forward of the canonical forms for them. For each cell we find that the resulting differential form has singularities corresponding to spurious boundaries between polytopes in a triangulation. For example, in the first positroid triangulations above, we find a singularity at $x_1+x_2=1$. However, this singularity disappears in the sum of terms, and we get a differential form in the so-called {\it local form}, with all singularities corresponding to the boundaries of the hypersimplex $\Delta_{2,4}$. We find the following explicit expression for $\omega_{2,4}$:
\begin{align}\label{eq:omega24}
    \omega_{2,4}
    &= \dd\log\left(\frac{x_2}{x_1}\right)\wedge\dd\log\left(\frac{x_3}{x_1}\right)\wedge\dd\log\left(\frac{x_4}{x_1}\right)
    -\dd\log\left(\frac{1-x_2}{x_1}\right)\wedge\dd\log\left(\frac{x_3}{x_1}\right)\wedge\dd\log\left(\frac{x_4}{x_1}\right)\nonumber\\\nonumber
    &-\dd\log\left(\frac{x_2}{x_1}\right)\wedge\dd\log\left(\frac{1-x_3}{x_1}\right)\wedge\dd\log\left(\frac{x_4}{x_1}\right)-\dd\log\left(\frac{x_2}{x_1}\right)\wedge\dd\log\left(\frac{x_3}{x_1}\right)\wedge\dd\log\left(\frac{1-x_4}{x_1}\right)\\&-\dd\log\left(\frac{x_2}{1-x_1}\right)\wedge\dd\log\left(\frac{x_3}{1-x_1}\right)\wedge\dd\log\left(\frac{x_4}{1-x_1}\right).
\end{align}
Interestingly, this expression can also be understood in a different way: each three-form in \eqref{eq:omega24} is a differential form of a three-dimensional simplex, where the boundaries of each simplex can be read off from the singularities of the form. Then \eqref{eq:omega24} suggests that the hypersimplex $\Delta_{2,4}$ can be obtained from the simplex with boundaries at $x_i=0$ after removing from it four simplices with boundaries $x_i=1$, $x_j=0$ for $j\neq i$, for $i=1,\ldots,4$. This is indeed a correct statement, as is illustrated in figure \ref{fig:Delta24}.
Notice that \eqref{eq:omega24} is not manifestly invariant under the parity symmetry $x_i \leftrightarrow (1-x_i)$, which we would expect to be true for $\Delta_{2,4}$. In particular, a parity conjugate version of \eqref{eq:omega24} is
\begin{align}\label{eq:omega24p}
  \omega_{2,4}  &=\dd\log\left(\frac{1-x_2}{1-x_1}\right)\wedge\dd\log\left(\frac{1-x_3}{1-x_1}\right)\wedge\dd\log\left(\frac{1-x_4}{1-x_1}\right)\nonumber\\\nonumber
    &-\dd\log\left(\frac{1-x_2}{1-x_1}\right)\wedge\dd\log\left(\frac{1-x_3}{1-x_1}\right)\wedge\dd\log\left(\frac{x_4}{1-x_1}\right)\\&-\dd\log\left(\frac{1-x_2}{1-x_1}\right)\wedge\dd\log\left(\frac{x_3}{1-x_1}\right)\wedge\dd\log\left(\frac{1-x_4}{1-x_1}\right)\nonumber\\\nonumber
    &-\dd\log\left(\frac{x_2}{1-x_1}\right)\wedge\dd\log\left(\frac{1-x_3}{1-x_1}\right)\wedge\dd\log\left(\frac{1-x_4}{x_1}\right)\\&-\dd\log\left(\frac{1-x_2}{x_1}\right)\wedge\dd\log\left(\frac{1-x_3}{x_1}\right)\wedge\dd\log\left(\frac{1-x_4}{x_1}\right).
\end{align}
However, using the constraint $x_1+x_2+x_3+x_4=2$, one can show that \eqref{eq:omega24} and \eqref{eq:omega24p} are the same. The formula \eqref{eq:omega24p} provides an alternative inclusion-exclusion triangulation for $\Delta_{2,4}$.

Our study of the hypersimplex $\Delta_{2,4}$ can be easily generalized to $\Delta_{k,n}$ for any $n$ and $k$. In all these cases we need to find a single positroid triangulation of the hypersimplex $\Delta_{k,n}$, and to use the algebraic moment map $\tilde\mu$ to calculate the push-forward of differential forms on Grassmannian positroid cells, summing over the triangulation. This allows us to find a general formula for the hypersimplex canonical form $\omega_{k,n}$. Our result has logarithmic singularities on all boundaries of the hypersimplex $\Delta_{k,n}$, which are of the form $\Delta_{k-1,n-1}$ or $\Delta_{k,n-1}$, and the residue when evaluated at these boundaries is $\omega_{k-1,n-1}$ and $\omega_{k,n-1}$, respectively.

Before writing down an explicit form for $\omega_{k,n}$, we need to introduce some notation which will allow us to write it in a concise way. Let us consider a $(d-1)$-dimensional geometry with exactly $2d$ boundaries of two types: boundaries at hyperplanes $a_i=0$, $i=1,\ldots,d$, and boundaries at hyperplanes $b_i=0$, $i=1,\ldots,d$. We know that a generic set of $d$ hyperplanes in a $(d-1)$-dimensional space defines a simplex. Let us take $J\in \binom{[d]}{l}$ and denote by $\Sigma_J$ the simplex bounded by hyperplanes $a_j=0$ for $j\in [d]\setminus J$ and $b_{j'}=0$ for $j'\in J$. The canonical differential form $\sigma_J$ for the simplex $\Sigma_J$ is then
 \begin{equation}\label{eq:simp-vol-form}
\sigma_J= \bigwedge_{j=2}^{d} \dd\log\left(\frac{\alpha_j}{\alpha_1}\right), \text{ where } \alpha_j= \begin{cases}
      a_j, & \text{for } j \not\in J\,, \\
      b_j, & \text{for } j \in J \,.
    \end{cases}
\end{equation}
The choice of $\alpha_1$ in the denominator is arbitrary, and any other $\alpha_j$ can be chosen at the cost of an overall factor $(-1)^{j+1}$. The simplex described above has $d-l$ facets of the form $a_i=0$, and $l$ facets of the form $b_i=0$. It will prove useful to define a sum of the forms $\sigma_J$ over all simplices with this distribution of facets:
\begin{equation}\label{eq:Snl-def}
\sigma_{l,d}\coloneqq \sum_{J \in\binom{[d]}{l}}\sigma_J\,.
\end{equation}
This sum over all simplices with a specific facet distribution enjoys useful properties. First of all, there is an inductive way to find $\sigma_{l,d}$ from $\sigma_{l,d-1}$ and $\sigma_{l-1,d-1}$:
\begin{equation}
\sigma_{l,d} = \sigma_{l,d-1} \wedge \dd\log\left(\frac{a_d}{a_j}\right)+ \sigma_{l-1,d-1}\wedge\dd\log\left(\frac{b_d}{b_j}\right),
\end{equation}
for any $j=1,\ldots,d-1$. From this it immediately follows that:
\begin{align}
&\underset{a_d=0}{\Res}\sigma_{l,d} = \sigma_{l,d-1}\,,\\
&\underset{b_d=0}{\Res}\sigma_{l,d} = \sigma_{l-1,d-1}\,.
\end{align}
More generally, we can take a residue for $a_j=0$ or $b_j=0$ for any $j=1,\ldots,d$, and obtain similar formulae with the right hand side relabelled.
Another identity we will use is: 
\begin{equation}\label{eq:Snl-id-one}
\sum_{l=0}^d (-1)^l \sigma_{l,d}=0\,.
\end{equation}
Also, let us notice that the parity symmetry that exchanges $a_j$ with $b_j$ leads to
\begin{equation}
\sigma_{l,d}\xleftrightarrow{a_j \leftrightarrow b_j} \sigma_{d-l,d}\,.
\end{equation}
Finally, by expanding $\dd\log\left(\alpha_j/\alpha_1\right) = \dd\log \alpha_j - \dd \log \alpha_1$ one can alternatively write \eqref{eq:simp-vol-form} as:
\begin{equation}\label{eq:simp-vol-form-alt}
\sigma_J = \sum_{j=1}^d (-1)^{j+1} \bigwedge_{i \in [d]\setminus\{j\}}\dd\log(\alpha_i),\quad \alpha_i= \begin{cases}
      a_i, & \text{if } i \not\in J\,, \\
      b_i, & \text{if } i \in J \,.
    \end{cases}
\end{equation}
Note that the $d$ terms in the sum of \eqref{eq:simp-vol-form-alt} can be divided into two categories: there are $l$ terms with $l$ one-forms $\dd\log b_i$'s and $d-l-1$ one-forms $\dd\log a_i$'s, and there are $d-l$ terms with $l-1$ one-forms $\dd\log b_i$'s and $d-l$ one-forms $\dd\log a_i$'s. We introduce the notation
\begin{equation}
\tau_{l,d}\coloneqq\sum_{j=1}^d (-1)^{j+1} \sum_{I\in\binom{[d]\setminus\{j\}}{l}}\bigwedge_{i\in[d]\setminus\{j\}}\dd\log\alpha_i,\quad \alpha_i= \begin{cases}
      a_i, & \text{if } i \not\in I\,, \\
      b_i, & \text{if } i \in I \,,
    \end{cases}
\end{equation}
which is the sum over all terms with exactly $l$ $\dd\log b_i$'s and $d-l-1$ $\dd \log a_i$'s with minus signs consistent with \eqref{eq:simp-vol-form-alt}. It then follows that:
\begin{equation}\label{eq:S-L-relation}
\sigma_{l,d} = \tau_{l,d}+\tau_{l-1,d}\,.
\end{equation}
This also provides a natural interpretation for equation \eqref{eq:Snl-id-one}, as the alternating sum makes the terms in \eqref{eq:S-L-relation} telescope, and we use the fact that $\tau_{-1,d}=\tau_{d,d}=0$.

Armed with this formalism we can now set $a_i=x_i$ and $b_i=1-x_i$, and write the canonical differential form $\omega_{k,n}$ for the hypersimplex $\Delta_{k,n}$ for general $n$ and $k$ as:
\begin{equation}\label{eq:hyp-form}
\omega_{k,n} = \sum_{l=0}^{k-1} (-1)^l \sigma_{l,n} = \sum_{l=0}^{n-k+1} (-1)^{n-l}\sigma_{n-l,n}\,.
\end{equation}
The equality between these two expressions comes from \eqref{eq:Snl-id-one} and the fact that on the support of the hypersimplex constraint $x_1+\ldots+x_n=k$ we have:
\begin{equation}
\sigma_{k,n}=0\,.
\end{equation}
As mentioned before, the alternating minus signs have the effect that terms telescope when expanded using \eqref{eq:S-L-relation}. This allows us to write the hypersimplex form as a single term:
\begin{equation}\label{eq:hyp-form-2}
\omega_{k,n} =\tau_{k-1,n}\,.
\end{equation}
Using the properties of the forms $\sigma$ and $\tau$, we can  immediately read off the following properties for the hypersimplex canonical forms:
\begin{align}
&\omega_{k,n}\xleftrightarrow{x_i \leftrightarrow 1-x_i} \omega_{n-k,n}\,,\\
&\underset{x_n=0}{\Res} \omega_{k,n} = \omega_{k,n-1}\,,\\
&\underset{x_n=1}{\Res}\omega_{k,n} = \omega_{k-1,n-1}\,.
\end{align}
This reflects the proper structure of hypersimplex boundaries, and the fact that $\Delta_{k,n}$ is parity dual to $\Delta_{n-k,n}$.

We summarize this section by rewriting the results we obtained above for $k=1$, $k=n-1$, and $n=4,k=2$ using this generalized notation, and by providing some additional simple examples. 
For the cases when the hypersimplex is a simplex, namely $k=1$ and $k=n-1$, we can write
\begin{align}
\omega_{1,n}&=\sigma_{0,n}=\sigma_{\emptyset}\,,\label{eq:mom-k=1-form}\\
\omega_{n-1,n}&=\sigma_{n,n}=\sigma_{[n]}\,.\label{eq:mom-k=n-1-form}
\end{align}
For $n=4$, $k=2$ we simply find
\begin{equation}
\omega_{2,4}=\sigma_{0,4}-\sigma_{1,4}=\sigma_{\emptyset}-\sigma_{\{1\}}-\sigma_{\{2\}}-\sigma_{\{3\}}-\sigma_{\{4\}}\,,
\end{equation}
where the second expression corresponds to the inclusion-exclusion triangulation of $\Delta_{2,4}$ which we discussed after formula \eqref{eq:omega24}. We can also see a similar type of triangulation for higher $n$, for example for $\Delta_{2,n}$ we find
\begin{equation}
\omega_{2,n}=\sigma_{0,n}-\sigma_{1,n}=\sigma_{\emptyset}-\sum_{i=1}^n\sigma_{\{i\}}\,,
\end{equation}
where each $\sigma_{\{i\}}$ corresponds to an $(n-1)$-dimensional simplex with one facet at $x_i=1$ and all other facets at $x_j=0$ for $j\neq i$. These triangulations generalize to any $k$ and we obtain an inclusion-exclusion type of triangulation of $\Delta_{k,n}$:
\begin{equation}\label{eq:hyptriangulation}
\Delta_{k,n}=\Sigma_\emptyset\setminus\left(\bigcup_{I_1\in \binom{[n]}{1}}\Sigma_{I_1}\setminus\left(\bigcup_{I_2\in \binom{[n]}{2}}\Sigma_{I_2}\setminus\left(\ldots \setminus\bigcup_{I_{k-1}\in \binom{[n]}{k-1}}\Sigma_{I_{k-1}}\right)\right)\right),
\end{equation}
which to our knowledge has not been previously known.


\section{Momentum amplituhedron}
\label{sec:mom}
The momentum amplituhedron $\mathcal{M}_{n,k}$ is a positive geometry introduced in \cite{Damgaard:2019ztj} to describe tree-level scattering amplitudes in $\mathcal{N}=4$ sYM in spinor helicity space. Its counterpart in momentum twistor space is the amplituhedron $\mathcal{A}_{n,k}$ \cite{Arkani-Hamed:2013jha}, which has a natural generalization $\mathcal{A}_{n,k}^{(m)}$ beyond the case relevant to physics, labelled by an integer $m$, with $m=4$ corresponding to the physical case. It was observed in \cite{Lukowski:2020dpn} that a natural generalization also exists for the momentum amplituhedron for even $m$, and the authors of \cite{Lukowski:2020dpn}, including one of the authors of this paper, suggested a possible definition for $\mathcal{M}_{n,k}^{(m)}$ for even $m$. In particular, they conjectured in section 12 of their paper that, for $m=2$, the momentum amplituhedron $\mathcal{M}_{n,k}^{(2)}$ shares many properties with the hypersimplex $\Delta_{k,n}$. Their main conjecture stated that the positroid triangulations of the hypersimplex $\Delta_{k,n}$ are in one-to-one correspondence with positroid triangulations of the momentum amplituhedron $\mathcal{M}_{n,k}^{(2)}$. Based on this, it was found in \cite{Lukowski:2020bya} that the boundary stratification of the momentum amplituhedron $\mathcal{M}_{n,k}^{(2)}$ is analogous to the boundary stratification of the hypersimplex $\Delta_{k,n}$. In this section we show that both statements are not correct and find their counterexamples. 

Despite the fact that the definition of $\mathcal{M}_{n,k}^{(2)}$ in \cite{Lukowski:2020dpn} does not provide an object with desired properties, we find interesting differential forms that can be naturally defined in the space introduced there. These differential forms have properties analogous to the hypersimplex canonical forms $\omega_{k,n}$ we studied in section \ref{sec:hypersimplex}. They are not, however, canonical forms of the momentum amplituhedron $\mathcal{M}_{n,k}^{(2)}$ defined in \cite{Lukowski:2020dpn}. 
 
\subsection{Definition of \texorpdfstring{$m=2$}{} momentum amplituhedron}
\label{sec:momamp}
We follow the notation in \cite{Lukowski:2020dpn} and provide the definition of the momentum amplituhedron $\mathcal{M}_{n,k}^{(2)}$. It relies on two matrices $\Lambda$ and $\widetilde\Lambda$, encoding the ``external data'':
\begin{equation}
\Lambda=(\Lambda_{1} \Lambda_{2} \ldots \Lambda_{n}) \in M(n-k+1, n), \quad \widetilde{\Lambda}=(\widetilde{\Lambda}_{1} \widetilde{\Lambda}_{2} \ldots \widetilde{\Lambda}_{n}) \in M(k+1, n)\,.
\end{equation}
One assumes that $\Lambda$ is a positive matrix, i.e. all its maximal minors are positive, and $\widetilde\Lambda$ is a twisted positive matrix, i.e. the matrix describing its orthogonal complement is a positive matrix. Then, the $m=2$ momentum amplituhedron $\mathcal{M}^{(2)}_{n,k}$ is defined as the image of the positive Grassmannian $G_+(k,n)$ through the map specified by these matrices:
\begin{align}\label{eq:PhiLLt}
\Phi_{(\Lambda, \widetilde{\Lambda})}: G_{+}(k, n)\rightarrow G\left(n-k, n-k+1\right) \times G\left(k, k+1\right),\qquad
C\mapsto (Y,\widetilde Y)\,,
\end{align}
where
\begin{equation}
Y_{\alpha}^{A}=c_{\alpha i}^{\perp} \Lambda_{i}^{A}\,, \quad \widetilde{Y}_{\dot{\alpha}}^{\dot{A}}=c_{\dot{\alpha} i} \widetilde{\Lambda}_{i}^{\dot{A}}\,.
\end{equation}
We use $C=\{c_{\dot\alpha i}\}\in G_+(k,n)$, and $C^{\perp}=\left\{c_{\alpha i}^{\perp}\right\}$ is the orthogonal complement of $C$. The image of the positive Grassmannian naturally lives in an $(n-1)$-dimensional subspace of the $(n-k+k=n)$-dimensional space $ G\left(n-k, n-k+1\right) \times G\left(k, k+1\right)$ specified by the `momentum conservation'-like identity:
\begin{equation}\label{eq:mom-cons}
\sum_{i=1}^n \left(Y^\perp \cdot\Lambda\right)_i \left(\widetilde Y^\perp \cdot\widetilde\Lambda\right)_i=0\,.
\end{equation}
Similar to the $m=4$ momentum amplituhedron $\mathcal{M}_{n,k}^{(4)}$ in \cite{Damgaard:2019ztj}, we define the `spinor helicity' variables $\lambda,\tilde\lambda$ as:
\begin{align}
\lambda_i&\coloneqq \left< Y i \right>= \epsilon_{A_1 A_2\cdots A_{n-k} A_{n-k+1}} Y_1^{A_1} Y_2^{A_2}\cdots Y_{n-k}^{A_{n-k}} \Lambda_i^{A_{n-k+1}}\,, \\ 
\tilde\lambda_i&\coloneqq [ \widetilde Y i ]= \epsilon_{\dot A_1 \dot A_2\cdots \dot A_{k}\dot A_{k+1}} \widetilde Y_1^{\dot A_1} \widetilde Y_2^{\dot A_2}\cdots \widetilde Y_{k}^{\dot A_{k}} \widetilde\Lambda_i^{\dot A_{k+1}}\,.
\end{align}
These $\lambda$ and $\tilde \lambda$ variables satisfy a similar `momentum conservation' identity:
\begin{equation}
\sum_{i=1}^n \lambda_i \tilde\lambda_i=0\,.
\end{equation}

\subsection{Momentum amplituhedron-like logarithmic forms}\label{sec:momamp-diff-form}

Before discussing the geometry of $\mathcal{M}_{n,k}^{(2)}$, let us focus on differential forms that can be defined in the $(\lambda,\tilde\lambda)$ space. Since the domain of the maps $\tilde\mu$ and $\Phi_{(\Lambda,\widetilde\Lambda)}$ are the same, a natural question is what happens when we take a collection of positroid cells in the positive Grassmannian $G_+(k,n)$ that provides a positroid triangulation of the hypersimplex $\Delta_{k,n}$, and evaluate their push-forwards using the momentum amplituhedron map $\Phi_{(\Lambda,\widetilde\Lambda)}$ \eqref{eq:PhiLLt}. An important observation is that this push-forward does not depend on the positivity conditions for the $\Lambda$ and $\widetilde\Lambda$ matrices.

Taking any collection of $G_+(k,n)$ positroid cells labels $\mathcal{T}=\{\pi_1,\ldots,\pi_p\}$ that gives a positroid triangulation of $\Delta_{k,n}$, we can define
\begin{equation}\label{eq:momampformdef}
\overline\omega_{n,k}=\sum_{\pi\in \mathcal{T}} (\Phi_{(\Lambda,\widetilde\Lambda)})_*\, \omega_\pi\,,
\end{equation}
where $\omega_\pi$ is the canonical form of the positroid cell $S_\pi$, and $(\Phi_{(\Lambda,\widetilde\Lambda)})_*$ indicates the push-forward\footnote{The signs of push-forwards are fixed such that the common singularities appearing in different terms, i.e.~the spurious singularities, have a vanishing residue. We found that it is always possible to find such combinations of signs.}.  We have calculated $\overline\omega_{n,k}$ using positroid triangulations of hypersimplices up to $n=7$, all $k$, and found that the answer is independent from the triangulation. Moreover, it can be expressed using the notation we introduced in section \ref{sec:hyp-form}. By taking $\sigma_{l,n}$ defined in \eqref{eq:simp-vol-form-alt}, and substituting $a\to \lambda,\,b\to\tilde\lambda$, we can write the differential form $\overline\omega_{n,k}$ in \eqref{eq:momampformdef} as:
\begin{align}\label{eq:mom-volume-form}
\overline\omega_{n,k} &= \sum_{l=0,2,4,\ldots}^{k-1} \sigma_{l,n}\,,\quad \text{for } k\text{ odd}\,,\\
\overline\omega_{n,k}& = \sum_{l=1,3,5,\ldots}^{k-1} \sigma_{l,n}\,,\quad \text{for } k\text{ even}\,.
\end{align}
We believe that these formulae are true for any $n$ and $k$. These can also be written in a more uniform way using the differential forms $\tau$ from \eqref{eq:simp-vol-form-alt} as
\begin{equation}
\overline\omega_{n,k}=\sum_{l=0}^{k-1} \tau_{l,n}\,.
\end{equation}
Interestingly, these differential forms have properties similar to those we have found for the hypersimplex canonical forms $\omega_{k,n}$. In particular, they are parity symmetric when $\lambda$ is exchanged with $\tilde\lambda$:
\begin{align}
&\overline\omega_{n,k}\xleftrightarrow{\lambda_i \leftrightarrow \tilde\lambda_i} \overline\omega_{n-k,k}\,.\label{eq:mom-form-parity}
\end{align}
This can be shown using a version of equation \eqref{eq:Snl-id-one}: 
\begin{equation}
\sum_{l=0,2,4,\ldots}^{\leq n}\sigma_{l,n}=\sum_{l=1,3,5,\ldots}^{\leq n}\sigma_{l,n}\,,
\end{equation}
and the fact that on the support of momentum conservation we have:
\begin{equation}
\sum_{l=0,2,4,\ldots}^{\leq n}\sigma_{l,n}=\sum_{l=1,3,5,\ldots}^{\leq n}\sigma_{l,n}=0\,,\qquad \text{for} \qquad \sum_{i=1}^n \lambda_i \tilde\lambda_i=0\,.
\end{equation}
Additionally, the differential form $\overline\omega_{n,k}$ has an identical singularity structure with the hypersimplex canonical forms $\omega_{k,n}$, namely:
\begin{align}\label{eq:momampres1}
&\underset{\lambda_n=0}{\text{Res}}\;\overline\omega_{n,k} = \overline\omega_{n-1,k}\,,\\\label{eq:momampres2}
&\underset{\tilde\lambda_n=0}{\text{Res}}\;\overline\omega_{n,k} = \overline\omega_{n-1,k-1}\,.
\end{align}
Analogous formulae are also true if we replace $\lambda_n, \tilde\lambda_n$ with any other $\lambda_i,\tilde\lambda_i$ for $i=1,2,\ldots,n$. These formulae indicate that the structure of singularities of the differential form $\overline\omega_{n,k}$ is exactly the same as the structure of singularities of $\omega_{k,n}$ in section \ref{sec:hyp-form}, after we identify $\lambda_i$ with $x_i$, and $\tilde\lambda_i$ with $1-x_i$. In particular, there are exactly $2n$ singularities, $n$ of which are of the form $\lambda_i=0$, and $n$ of which are of the form $\tilde\lambda_i=0$. The residues at these singularities are given by differential forms $\overline\omega$ with lower labels as in \eqref{eq:momampres1} and \eqref{eq:momampres2}, providing us with a recursive description akin to the one for the hypersimplex canonical forms $\omega_{k,n}$.

\subsection{Geometry}

Our calculations in the previous section pose a natural question whether there exists a geometric object for which $\overline\omega_{n,k}$ provides the canonical differential form. The first guess would be that this object must be the momentum amplituhedron defined in section \ref{sec:momamp}. We have however checked that even in the first non-trivial example, for $n=4, k=2$, the momentum amplituhedron $\mathcal{M}_{4,2}^{(2)}$ defined above is not the correct geometry. Instead, one needs to modify the positivity conditions in the definition of $\mathcal{M}_{4,2}^{(2)}$ to get a geometry with $\overline{\omega}_{4,2}$ as the canonical form. Even after this modification, the final conjecture of section 12 in \cite{Lukowski:2020dpn} is still not correct since, depending on the choice of external data, only one out of two positroid triangulations of the hypersimplex $\Delta_{2,4}$ provides a triangulation of such modified momentum amplituhedron. This can be attributed to the fact that, even with the modified positivity conditions, the region we define is concave. We have found that for $k=2$ and any $n$ we can always find conditions for external data $\Lambda$ and $\widetilde{\Lambda}$ such that the resulting geometry can be triangulated using some, but not all, of the positroid triangulations of the hypersimplex $\Delta_{2,n}$. Similar statement holds true for $k=n-2$, as well as for $n=6$ and $k=3$. It is, however, not possible beyond these cases and therefore we conclude that $\Phi_{(\Lambda,\widetilde\Lambda)}$ cannot be used to define a geometry for which $\overline\omega_{n,k}$ is the canonical differential.

Let us start by stating that for $k=1$ and for $k=n-1$ the momentum amplituhedron $\mathcal{M}^{(2)}_{n,k}$ is just a simplex. In these cases, the map $\Phi_{(\Lambda,\widetilde\Lambda)}$ is injective and there is no need for any triangulation. Then the canonical differential form for $\mathcal{M}^{(2)}_{n,1}$ is
\begin{equation}
\overline\omega_{n,1}=\dd\log\left(\frac{\lambda_2}{\lambda_1}\right)\wedge\dots\wedge\dd\log\left(\frac{\lambda_n}{\lambda_1}\right),
\end{equation}
and for $\mathcal{M}^{(2)}_{n,n-1}$ is
\begin{equation}
\overline\omega_{n,n-1}=\dd\log\left(\frac{\tilde\lambda_2}{\tilde\lambda_1}\right)\wedge\dots\wedge\dd\log\left(\frac{\tilde\lambda_n}{\tilde\lambda_1}\right),
\end{equation}   
Trivially, the boundary stratifications of $\mathcal{M}^{(2)}_{n,1}$ and $\mathcal{M}^{(2)}_{n,n-1}$ are equivalent to the boundary stratifications of the hypersimplices $\Delta_{1,n}$ and $\Delta_{n-1,n}$, respectively.

\begin{figure}
\begin{center}
\includegraphics[scale=0.5]{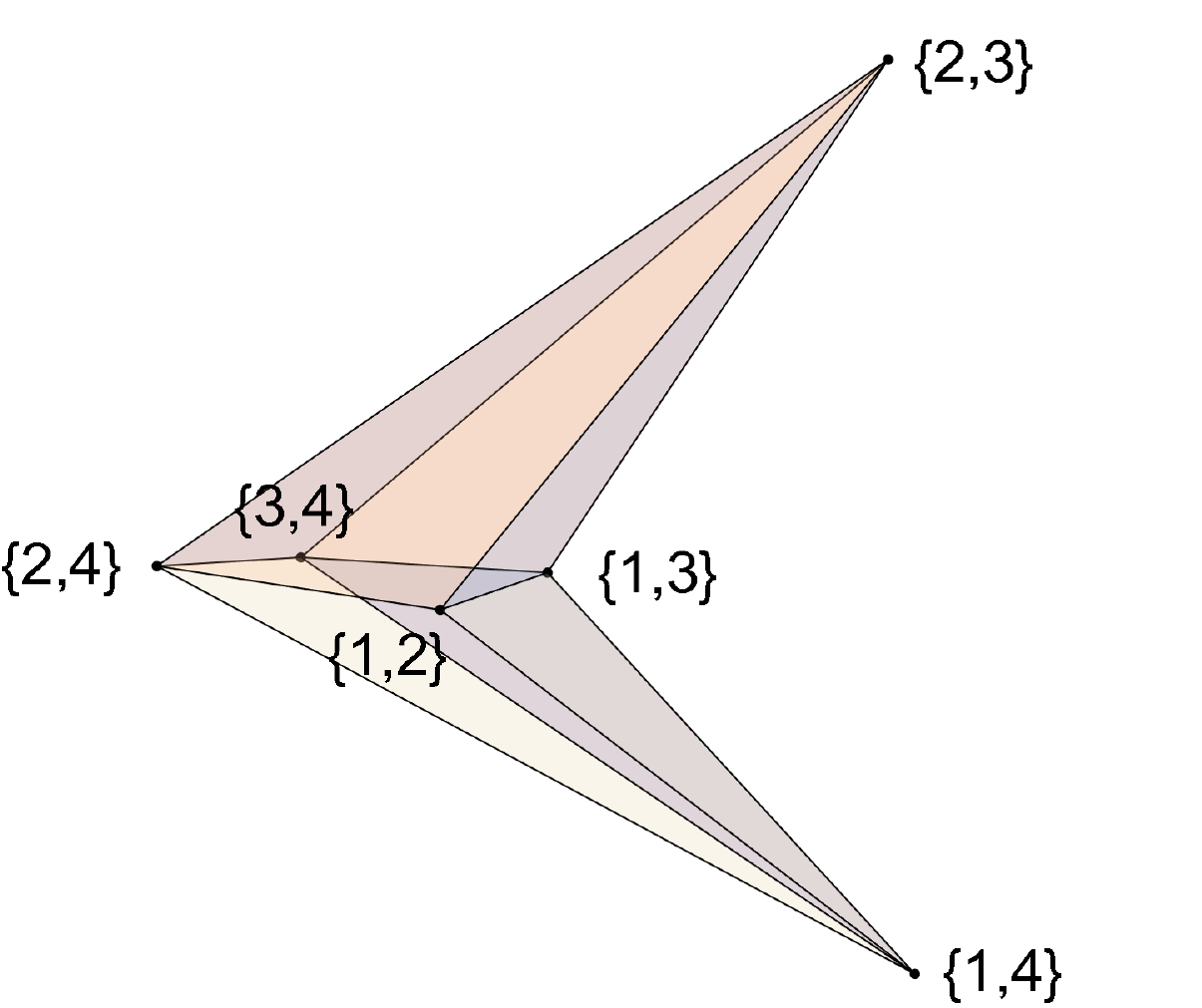}
\end{center}
\caption{Image of the positive Grassmannian $G_+(2,4)$ through the map $\Phi_{(\Lambda,\widetilde\Lambda)}$ with positivity conditions \eqref{eq:otherpositivity}.}
\label{fig:M42}
\end{figure}

Beyond $k=1$ and $k=n-1$, the map $\Phi_{(\Lambda,\widetilde\Lambda)}$ is not injective anymore, as was the case for the algebraic moment map $\tilde\mu$. There are, however, significant differences between the two geometries that we illustrate in detail in the simplest non-trivial case: $n=4,k=2$. Recall that the hypersimplex $\Delta_{2,4}$ is a octahedron depicted in Fig.~\ref{fig:Delta24}, and it can be subdivided using pairs of positroid polytopes in two different ways. These positroid polytopes are images of 3-dimensional cells in the positive Grassmannian $G_+(2,4)$ through the algebraic moment map $\tilde\mu$. In particular, they have spurious boundaries along the hyperplanes $x_1+x_2=1$ or $x_2+x_3=1$. A similar analysis can be done using the $\Phi_{(\Lambda,\widetilde\Lambda)}$ map and the images of the three-dimensional cells have boundaries along $\lambda_1\tilde\lambda_1+\lambda_2\tilde\lambda_2=0$ (two cells) or $\lambda_2\tilde\lambda_2+\lambda_3\tilde\lambda_3=0$ (two cells). For a pair of cells to be a triangulation of $\mathcal{M}_{4,2}^{(2)}$, their images need to sit on the opposite sides of these spurious boundaries. This provides restrictions on the matrices $\Lambda$ and $\widetilde\Lambda$.    For example, for the cell parametrized by permutation $\{3,4,6,5\}$ we find
\begin{equation}\label{eq:l2l3}
\lambda_2\tilde\lambda_2+\lambda_3\tilde\lambda_3=([123]-\alpha_1[234])(-\langle 134\rangle+\alpha_2 \langle 124\rangle)\alpha_2\alpha_3\,,
\end{equation}
with $\alpha_i>0$, while for the cell parametrized by the permutation $\{4,3,5,6\}$ we find
\begin{equation}\label{eq:l2l3p}
\lambda_2\tilde\lambda_2+\lambda_3\tilde\lambda_3=-(\langle 234\rangle+\beta_1\beta_2\langle 123\rangle)([124]+\beta_3[134])\beta_1\,,
\end{equation}
with $\beta_i>0$. The surface $\lambda_2\tilde\lambda_2+\lambda_3\tilde\lambda_3=0$ is the shared boundary of these images, and to have a triangulation we need to enforce a uniform, and opposite, sign of expressions \eqref{eq:l2l3} and \eqref{eq:l2l3p} for all $\alpha_i>0$ and $\beta_i>0$. It is easy to check that this is not the case if we assume the positivity conditions from section \ref{sec:momamp}, providing a counter-example to the statements in section 12 of \cite{Lukowski:2020dpn}. Instead, we should take for example 
\begin{equation}\label{eq:otherpositivity}
[123]>0,[124]>0,[134]>0,[234]<0,\qquad \langle 123\rangle>0\,,\langle 124\rangle>0,\langle 134\rangle<0,\langle 234\rangle>0\,.
\end{equation} 
With these conditions, the $\Phi_{(\Lambda,\widetilde\Lambda)}$ images of cells labelled by permutations $\pi_1=\{3,4,6,5\}$ and $\pi_2=\{4,3,5,6\}$ subdivide the image of the positive Grassmannian $G_+(2,4)$, and therefore the logarithmic form $\overline{\omega}_{2,4}$ is the canonical form of this geometry. However, in this case, the images of the remaining two cells, $\pi_3=\{2,4,5,7\}$ and $\pi_4=\{3,5,4,6\}$, do overlap and they do not provide a subdivision of $\mathcal{M}_{4,2}^{(2)}$. This comes from the fact that the image of the positive Grassmannian $G_+(2,4)$ through $\Phi_{(\Lambda,\widetilde\Lambda)}$ with the positivity conditions \eqref{eq:otherpositivity} is concave and looks like the shape in figure \ref{fig:M42}. 

For higher $n$, this becomes even more involved. For example, for $\mathcal{M}_{n,2}^{(2)}$ with $n>5$, there exist collections of cells in $G_+(2,n)$ that form a positroid triangulation of $\Delta_{2,n}$, but for which there are no matrices $\Lambda$ and $\widetilde\Lambda$ such that the images of these cells through $\Phi_{(\Lambda,\widetilde\Lambda)}$ are disjoint. In our investigations, we have found that for $n>4$ and $k=2$ there exist exactly $n$ positroid triangulations of the hypersimplex $\Delta_{2,n}$ for which positivity conditions for $\Lambda$ and $\widetilde\Lambda$ can be found to render a triangulation. In all these cases, the differential form $\overline\omega_{n,2}$ from \eqref{eq:mom-volume-form} is the canonical differential form of the corresponding image of $G_+(2,n)$ through $\Phi_{(\Lambda,\widetilde\Lambda)}$. Even this becomes impossible for higher $k$: we found that for $n>6$ and $2<k<n-2$ there are no triangulations of $\Delta_{k,n}$ for which the images through the map $\Phi_{(\Lambda,\widetilde\Lambda)}$ are disjoint. This shows that one cannot use the map  $\Phi_{(\Lambda,\widetilde\Lambda)}$ to generate a region in the $(\lambda,\tilde\lambda)$-space for which $\overline\omega_{n,k}$ is the canonical form.
\section{Summary and outlook}
In this paper, we have studied two geometries, the hypersimplex $\Delta_{k,n}$ and the generalization of the momentum amplituhedron $\mathcal{M}_{n,k}^{(2)}$ proposed in \cite{Lukowski:2020dpn}, from the point of view of positive geometries. We have provided two main results. One is the previously unknown formula \eqref{eq:hyp-form} for the hypersimplex canonical form $\omega_{k,n}$. The formula has a natural interpretation as a new inclusion-exclusion triangulation of hypersimplex given in \eqref{eq:hyptriangulation}. Moreover, we provide a negative but important result stating that the generalization of the momentum amplituhedron suggested in \cite{Lukowski:2020dpn} does not possess the desired properties. In particular, we have found counter-examples showing that the conjectures in section 12 of \cite{Lukowski:2020dpn} regarding positroid triangulations of $\mathcal{M}_{n,k}^{(2)}$ are not valid. It can be attributed to the fact that the momentum amplituhedron for $m=2$ is ``concave''. This, in turn, is related to the fact that the momentum amplituhedron for $m=2$ shares properties with the ordinary amplituhedron for $m=1$. The latter is known to be concave and, in general, amplituhedra for odd $m$ are less well-behaving than the ones for even $m$, see for example \cite{Ferro:2018vpf}. We predict that the momentum amplituhedron for $m=2,6,10,\ldots$ will have similar behaviour, and the conjectures from section 12 of \cite{Lukowski:2020dpn} will not hold in these cases. The question remains open on whether the conjectures are correct for $m$ divisible by four, beyond $m=4$. 

In this paper we have also provided interesting differential forms written directly in the $(\lambda,\tilde\lambda)$ spinor helicity space, which have properties analogous to those of the hypersimplex canonical forms. This leads to the question of whether one can find a shape inside the $(\lambda,\tilde\lambda)$ space with the canonical differential form given by $\overline\omega_{n,k}$. It is  unclear from our explorations whether it will be possible, and it remains an interesting open problem.

\section*{Acknowledgements}

We would like to thank Livia Ferro and Lauren Williams for useful discussions.


\appendix

\section{Definition of positive geometry and push-forward}
\label{app:positive}
Positive geometries \cite{Arkani-Hamed:2017tmz} naturally live in complex projective spaces $\mathbb{P}^N$, and their real parts $\mathbb{P}^N(\mathbb{R})$. One defines $X$ to be a complex projective algebraic variety of complex dimension $D$ and $X(\mathbb{R})$ to be its real part, and one denotes by $X_{\geq 0} \subset X(\mathbb{R})$ an oriented set of real dimension $D$.
 A $D$-dimensional {\it positive geometry} is a pair $(X, X_{\geq 0})$ equipped with a unique non-zero differential $D$-form $\Omega(X, X_{\geq 0})$, called the {\it canonical form}, satisfying the following recursive axioms:
\begin{itemize}
\item For $D = 0$ we have  that $X=X_{\geq 0}$ is a single real point and $\Omega(X, X_{\geq 0})=\pm 1$ depending on the orientation of $X_{\geq 0}$.
\item  For $D > 0$ we have that every boundary component $(C, C_{\geq 0})$ of $(X, X_{\geq 0})$ is a positive geometry of dimension $D-1$. Moreover, the form $\Omega(X, X_{\geq 0})$ is constrained by the residue relation 
\begin{equation}
\mbox{Res}_C\, \Omega(X, X_{\geq 0}) = \Omega(C, C_{\geq 0})\,,
\end{equation}
 along every boundary component $C$, and has no singularities elsewhere.
\end{itemize}
The {\it residue} operation $\mbox{Res}_C$ for a meromorphic form $\omega$ on $X$ is defined in the following way: suppose $C$ is a subvariety of $X$ and $z$ is a holomorphic coordinate whose zero set $z = 0$ parametrizes $C$. Denote as $u$ the remaining holomorphic coordinates. Then a simple pole of $\omega$ at $C$ is a singularity of the form
\begin{equation}
\omega(u, z) = \omega'(u) \wedge \frac{dz}{z}
+ \ldots\,,
\end{equation}
where the ellipsis denotes terms smooth in the small $z$ limit, and $\omega'(u)$ is a non-zero meromorphic form on the boundary component. One defines
\begin{equation}
\text{Res}_C\, \omega := \omega'\,.
\end{equation}
If there is no such simple pole then one defines the residue to be zero.

We also define what we mean by the push-forward of a differential form. We consider a surjective meromorphic map  $\phi :A \to B$ of finite degree $p$, where $A$ and $B$ are complex manifolds of the same dimension. For a given point $b\in B$ we can find its pre-image, namely a collection of points $a_i$ in $A$, $i=1,\ldots,p$, satisfying $\phi(a_i)=b$. Taking a neighbourhood $U_i$ of each point $a_i$ and a neighbourhood $V$ of $b$, we can define the inverse maps: $\psi_i=\phi|_{U_i}^{-1}:V\to U_i$. Then the push-forward of a meromorphic top form $\alpha$ on $A$ through $\phi$ is a differential form $\beta$ on $B$ given by the sum over all solutions of the pull-backs through the inverse maps $\psi_i$:
\begin{equation}
\beta=\phi_*\alpha=\sum_{i=1}^p \psi_i^*\alpha \,,
\end{equation}
where the pull-back of a differential form is a standard notion in differential geometry.
In practice, one solves the equation $y=\phi(x)$ and for each solution $x=\psi_i(y)$ one substitutes the explicit expression for $x$ into the differential form $\alpha$, and then sums the resulting forms.

\bibliographystyle{nb}

\bibliography{mom_amp_vs_hyper}

\end{document}